\documentclass{pasa}
\usepackage{graphicx}
\usepackage{amsmath}
\usepackage[authoryear]{natbib}

\title[Magnification of a Point-Source-Point-Lens System]
  {A Note on the Overall Magnification of a Gravitational Point-Source-Point-Lens System}
\author[Walters \& Forbes]
  {S.J. Walters and L.K. Forbes \\
  School of Mathematics and Physics, University of Tasmania, P.O. Box 37, Hobart, 7001, Tasmania, Australia}
\date{Released 2015 Xxxxx XX}

\begin{document}

\begin{abstract}
The total magnification due to a point lens has been of particular interest as the theorem that gravitational lensing results in light amplification for all observers appears to contradict the conservation of photon number. This has been discussed several times, and various resolutions have been offered. In this note, we use a kinematic approach to provide a formula for the magnification factor for the primary image accurate to first order and valid for rays leaving the source at any trajectory. We thus determine the magnification over a sphere surrounding the system. A new result found is that while the magnification dips below unity far from the optical axis as noted by others, it returns to unity directly behind the source.
\end{abstract}

\maketitle


\section{Introduction} \label{intro}
The magnification formula for the increase of light from a distant source when `lensed' by an intervening massive object, was given eighty years ago in a small article by \citet{ein}. The derivation of this formula was later clarified by \citet{refs} and others. An interesting feature of the formula is that the magnification factor is arbitrarily large near the optical axis (the line on which the source and lens are situated), and falls away as the observer moves far from that line, but never falls below unity. This may appear to create a paradox in that if the magnification is always greater than one, there must be an increase in photon number due to the presence of the lensing object, but a lensing object such as a black hole does not create photons. For discussions of this paradox see for example, \citet{sch} and \citet{jar}. We note that Einstein and Refsdal make no mention of a paradox. Rather, they were only considering the region near the optical axis and so developed this formula as an appropriate approximation for that region only.

More recently, \citet{wuck} has shown that the magnification factor does in fact drop below unity for observers far from the optical axis, and that this reduction in intensity balances the increase in intensity for observers near to the optical axis. That is, Wucknitz shows that at a certain angle from the optical axis, the magnification factor drops below one, and continues to decrease as the angle is increased, always remaining less than one. In this note, we use a kinematical approach to calculate the magnification factor. The results here are largely in agreement with the results of \citet{wuck}. However, we also show that very far from the optical axis, the magnification factor reaches a minimum, and then returns to unity as the observer moves to the point (on a sphere surrounding the source) furthest from the lens.

For clarity, we will briefly discuss the traditional thin lens approach and explicitly state the full formula for the magnification factor, from which the small angle approximation may be derived. Here we only consider the primary image, as we are concerned with magnification far from the optical axis, where secondary images are very weak (less than first order). In the approach detailed by Refsdal, we imagine two planes, normal to the optical axis, the plane containing the lensing object and the plane containing the observer. Light rays from the source to the observer are treated as two line segments, one from the source to the lens plane, and one from the lens plane to the observer's plane. The deflection angle by which these two segments differ is given by Einstein's deflection angle ($\beta$, described below). It can be shown that the magnification on the observer's plane, determined by comparing the infinitesimal solid angle $\text{d}\Omega$ at the source with that at the observer's plane is given by:

\begin{eqnarray}
M&=&\text{d}\Omega_s/\text{d}\Omega_o \nonumber \\
&=&\frac{L\sin \phi \big [U^2-2 U V \sin \phi +V^2\big ]^{3/2}}{\big[U \sin \phi -V \big ] \big [L_1 U^2+L_2 L^2 (\cos ^2 \phi+\beta \cot \phi)\big ]} \nonumber
\end{eqnarray}
where
\begin{eqnarray}
U&=&L\cos(\beta-\phi) \nonumber \\
V&=&L_2\sin\beta.\nonumber 
\end{eqnarray}
The constants $L_1$ and $L_2$ are, respectively, the distances from the lens plane to the source and to the observer, and $L=L_1+L_2$. The angle $\phi$ is the ray's initial trajectory, that is, angle from the optical axis when leaving the source, and $\beta=2 r_s/(L_1 \tan \phi)$ is the deflection of the ray when crossing the lens plane. The constant $r_s=2 m G/c^2$ is the Schwarzschild radius (where $m$ is the mass of the lensing object, and $G$ and $c$ are Newton's constant and the speed of light respectively). In this note we will use geometrized units, that is $G=c=1$. Assuming $\phi$ and $\beta$ to be small gives the traditional magnification for the primary image, which is always greater than one. Additionally, we can see that the first bracketed term in the denominator can go to zero, which corresponds to the `Einstein angle', the angle at which an observer co-linear with source and lens mass will see the source as a ring around the optical axis. This can be shown to correspond to a final angle from the optical axis of zero, as expected. That is, the infinite magnification produced by the simple point-source-point-lens (PSPL) model occurs on the optical axis. Equating the bracketed term to zero, and again assuming $\phi$ and $\beta$ to be small gives the standard value for this `Einstein angle' (see, for example \citet{refsur}, p. 132)

\begin{equation}
\phi_E=\sqrt{\frac{2 r_s L_2}{L_1 L}}.  \nonumber
\end{equation}
 
\begin{figure}
\vspace{1cm}
\caption{Magnification factor plotted against position on the observer's plane. Specific values of the three constants have been chosen as follows: $L_1=10, L_2=15, r_s=0.2$. A horizontal line has been drawn where the image is neither magnified nor de-magnified ($M=1$). The curve passing through the horizontal line is plotted using the full magnification formula. The curve staying above the line uses the small angle approximation.}
\includegraphics[width=\columnwidth]{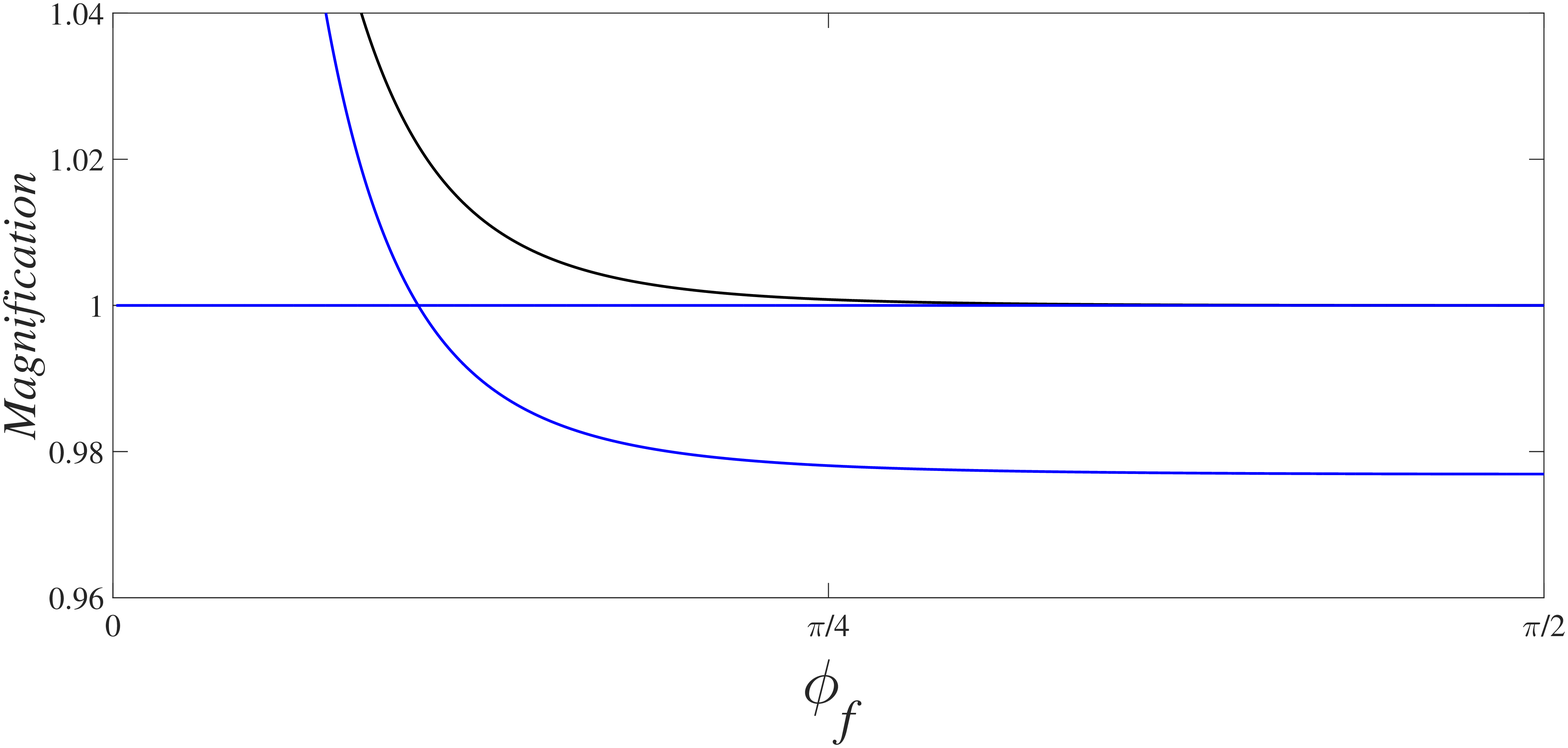}
\label{fig1}
\end{figure}

Parametrically plotting the magnification factor $M$ against $\phi_f$ as in Fig. \ref{fig1}, for some specific values of $r_s, L_1$ and $L_2$, we can see the magnification goes from infinite (when the first bracketed term in the denominator goes to zero, corresponding to the Einstein angle) down to unity for some value of $\phi_f$ and then remains below one out to $\pi/2$. For comparison, the small-angle approximation is plotted in the same figure (that is, assuming $\phi$ and $\beta$ small). As $\phi$ is increased to $\pi/2$, the limiting value for the magnification is

\begin{eqnarray}
M_{\pi/2}&=&1-\frac{2 r_s L_2}{L(L_1+2 r_s)} \nonumber \\
&=&1-\phi_E^2+O(r_s^2)  \nonumber
\end{eqnarray}

The value of $\phi$ where the magnification is equal to one has been approximated by \citet{wuck} to be the square root of the Einstein angle. That is, if we call this angle $\phi_{unity}$,

\begin{eqnarray}
\phi_{unity}&=&\sqrt[4]{\frac{2 r_s L_2}{L L_1}} \nonumber \\
&=&\sqrt{\phi_E}.  \nonumber
\end{eqnarray}

Re-arrangement of this equation allows us to replace $r_s$ by $L L_1 \phi^4/(2 L_2)$ in the full equation for the magnification, and confirm that at $\phi=\phi_{unity}$ the magnification $M=1+O(\phi^6)$.

In order to find the total magnification due to the PSPL system, we follow the approach of \citet{wuck}, by extending the magnification map from an observer's plane to a sphere surrounding the source. We will use kinematical ray equations, accurate to first order in the appropriate small parameter to derive the magnification on the surface of the sphere due to the single lensing mass.

\section{Kinematical approach}
In addition to being a poor approximation at large angles, the thin-lens approximation can account for only half of the rays leaving the source (the `forward' rays). In order also to account for the `backward' rays, we will now apply a fully consistent first order approximation to the question of overall flux due to gravitational lensing. Following the approach given in \citet{wf2}, we can identify the path equations for a ray leaving the origin at $\tau=0$ in a system containing a single massive lensing object with Schwarzschild radius $r_s$ located at $(x_m,y_m,z_m)$. The ray has initial trajectory with azimuthal angle $\phi$, and angle of inclination $\theta$ above the $x-y$ plane. To first order in $r_s$ the trajectory ($x,y,z$) of the light ray, parameterized using `time' $\tau$, thus becomes
\begin{eqnarray}
x&=&C_1\tau+\frac{r_s}{2}\bigg[\frac{C_1\tau}{R_0}+x_m P-Q \frac{x_m + C_1 B_m}{K_m}\bigg] \nonumber \\
y&=&C_3\tau+\frac{r_s}{2}\bigg[\frac{C_3\tau}{R_0}+y_m P-Q \frac{y_m + C_3 B_m}{K_m}\bigg] \nonumber \\
z&=&C_5\tau+\frac{r_s}{2}\bigg[\frac{C_5\tau}{R_0}+z_m P- Q\frac{z_m + C_5 B_m}{K_m}\bigg]
\label{xyz}
\end{eqnarray}
where
\begin{eqnarray}
R_0&=&\sqrt{\tau^2 + R_m^2 + 2 B_m \tau} \nonumber \\
R_m&=&\sqrt{x_m^2+y_m^2+z_m^2} \nonumber \\
B_m&=&-(C_1 x_m+C_3 y_m+C_5 z_m) \nonumber \\
K_m&=&R_m^2-B_m^2 \nonumber \\
P&=&\frac{1}{R_m}-\frac{1}{R_0} \nonumber \\
Q&=& \frac{B_m \tau}{R_m}(3-\frac{B_m^2}{R_m^2})+2 R_m-2 R_0, \nonumber
\label{ra}
\end{eqnarray}
and
\begin{eqnarray}
C_{1}&=&\cos \phi \cos \theta \nonumber \\
C_{3}&=&\sin \phi \cos \theta \nonumber \\
C_{5}&=&\sin \theta. \nonumber
\end{eqnarray}
Details are given in \citet{wf2}.
We now have the first order path equations for $x, y$ and $z$. In order to evaluate the flux of light rays at some outer sphere, we specify that the ray meets a sphere of radius $R$ at some `time' $\tau_f$. This enables us to solve for $\tau_f$ to first order in $r_s$. That is, let $\tau_f=T_0+r_s T_1+O(r_s^2)$, and solve the following equation for $T_0$ and $T_1$:
\begin{eqnarray}
R^2&=&x^2+y^2+z^2+O(r_s^2). \nonumber
\label{rsqr}
\end{eqnarray}
Solving this equation at zeroth and first order in $r_s$ gives $\tau_f$ to first order:
\begin{eqnarray}
T_0&=&R \nonumber \\
T_1&=&\frac{B_m}{R_m}-\frac{R + B_m}{R_f}, \nonumber
\label{t1}
\end{eqnarray}
where $R_f=\sqrt{R^2+R_m^2+2 B_m R}$. Substituting this value of $\tau$ into the path equations (\ref{xyz}), the values of $x, y$ and $z$ at the sphere can be calculated, given an initial trajectory ($\phi, \theta$). After some simplification, these may be written:
\begin{eqnarray}
x_f&=&C_1 R+\frac{r_s}{2}(x_a + C_1 B_a) F+O(r_s^2) \nonumber \\
y_f&=&C_3 R+\frac{r_s}{2}(y_a + C_3 B_a) F+O(r_s^2) \nonumber \\
z_f&=&C_5 R+\frac{r_s}{2}(z_a + C_5 B_a) F+O(r_s^2), \nonumber
\label{xyzf}
\end{eqnarray}
where the function $F$ is defined as
\begin{eqnarray}
F&=&\frac{1}{R_m}-\frac{1}{R_f} -2\frac{R_m-R_f}{K_m}-\frac{R B_m}{K_m R_m}(3 - \frac{B_m^2}{R_m^2}). \nonumber
\label{f}
\end{eqnarray}
These positions on the observer's sphere can be expressed in terms of azimuthal and inclination angles in the normal way, and may be simplified as:
\begin{eqnarray}
\phi_f&=&\arctan\frac{y_f}{x_f} \nonumber \\
&=&\phi+\frac{r_s}{2 R \cos^2\theta} (C_1 y_m - C_3 x_m) F +O(r_s^2)\nonumber \\
\theta_f&=&\arcsin\frac{z_f}{R} \nonumber \\
&=&\theta+\frac{r_s}{2 R \cos\theta}(z_m + C_5 B_m)F+O(r_s^2).
\label{phithetaf}
\end{eqnarray}
These equations define the intersection with the observer's sphere of a ray in a system with a lens located at $(x_m,y_m,z_m)$. In considering the problem of total flux for a system with a single lens, we may now orient the axes such that the lens object lies on the $x$-axis. Thus we set $y_m=z_m=0$. This will simplify equations (\ref{phithetaf}) to 
\begin{eqnarray}
\phi_f&=&\phi+\frac{r_s}{2 R \cos^2\theta} ( - C_3 x_m) F +O(r_s^2)\nonumber \\
\theta_f&=&\theta+\frac{r_s}{2 R \cos\theta}(-C_1 C_5 x_m)F+O(r_s^2). \nonumber
\end{eqnarray}
where $F$ and $R_f$ are now
\begin{eqnarray}
F&=&\frac{1}{x_m}-\frac{1}{R_f} +\frac{C_1 R}{x_m^2}+2\frac{R_f+C_1 R-x_m}{x_m^2(1-C_1^2)} \nonumber \\
R_f&=&\sqrt{R^2-2 x_m C_1 R+ x_m^2}. \nonumber
\end{eqnarray}
As in the introduction above, the magnification can be described by the ratio of the solid angles of a light bundle at the source and at the observer's sphere, such that:
\begin{eqnarray}
\frac{1}{M}&=&\frac{\partial \theta_f}{\partial \theta} \frac{\partial \phi_f}{\partial \phi} \frac{\cos \theta_f}{\cos \theta}
\label{M3}
\end{eqnarray}

In general the expression in equation (\ref{M3}) has a large number of terms. However, due to the $x$-axis symmetry of the current expression, the magnification is only a function of a single variable, that is the angle between the ray and the $x$-axis, represented by $C_1$. The magnification function will be the same for any of the great circles which passes through the $x$-axis. It is convenient then to choose the great circle lying in the $x-y$ plane. In that case, $C_5=0$ so that $\cos \theta_f=\cos \theta$, and (to first order in $r_s$) the other two components are

\begin{eqnarray}
\frac{\partial \theta_f}{\partial \theta}=1 + \frac{C_1 r_s}{2}\bigg(&-&\frac{2 R}{x_m(R_f - C_1 R + x_m)}\nonumber \\
&+& \frac{x_m}{R_f R} -\frac{C_1}{x_m}-\frac{1}{R}\bigg) \nonumber
\end{eqnarray}
\begin{eqnarray}
\frac{\partial \phi_f}{\partial \phi}=1 + \frac{r_s}{2}\bigg(&-&\frac{x_m^2(1-C_1^2)}{R_f^3}-\frac{2}{R_f}+\frac{C_1 x_m}{R_f R}+\frac{3}{x_m}\nonumber \\
&-&\frac{C_1}{R}-\frac{2 C_1^2}{x_m}+\frac{2 C_1 R}{x_m(R_f - C_1 R + x_m)}\bigg), \nonumber
\end{eqnarray}
where $C_1=\cos\phi$. The magnification is then
\begin{eqnarray}
M&=&1/\bigg(\frac{\partial \theta_f}{\partial \theta} \frac{\partial \phi_f}{\partial \phi}\bigg).
\label{M4}
\end{eqnarray}
Note that for an initial trajectory directly away from the lens mass $\phi=\pi$, that is $C_1=-1$, both the terms above become equal to one, giving no change in magnification.

\begin{figure}
\vspace{1cm}
\caption{Magnification factor plotted against $x$-position on the observer's sphere of radius $200$, centred on a source at the origin with a single lens with $r_s=0.5$ located on the positive $x$-axis at $x=20$. For the point-source-point-lens system it may be observed that the magnification is equal to unity directly behind the source, stays below one for most observer positions, but rises to infinity as observers approach the positive $x$-axis. The magnification value near the optical axis has been truncated to a maximum of $1.01$.}
\includegraphics[width=\columnwidth]{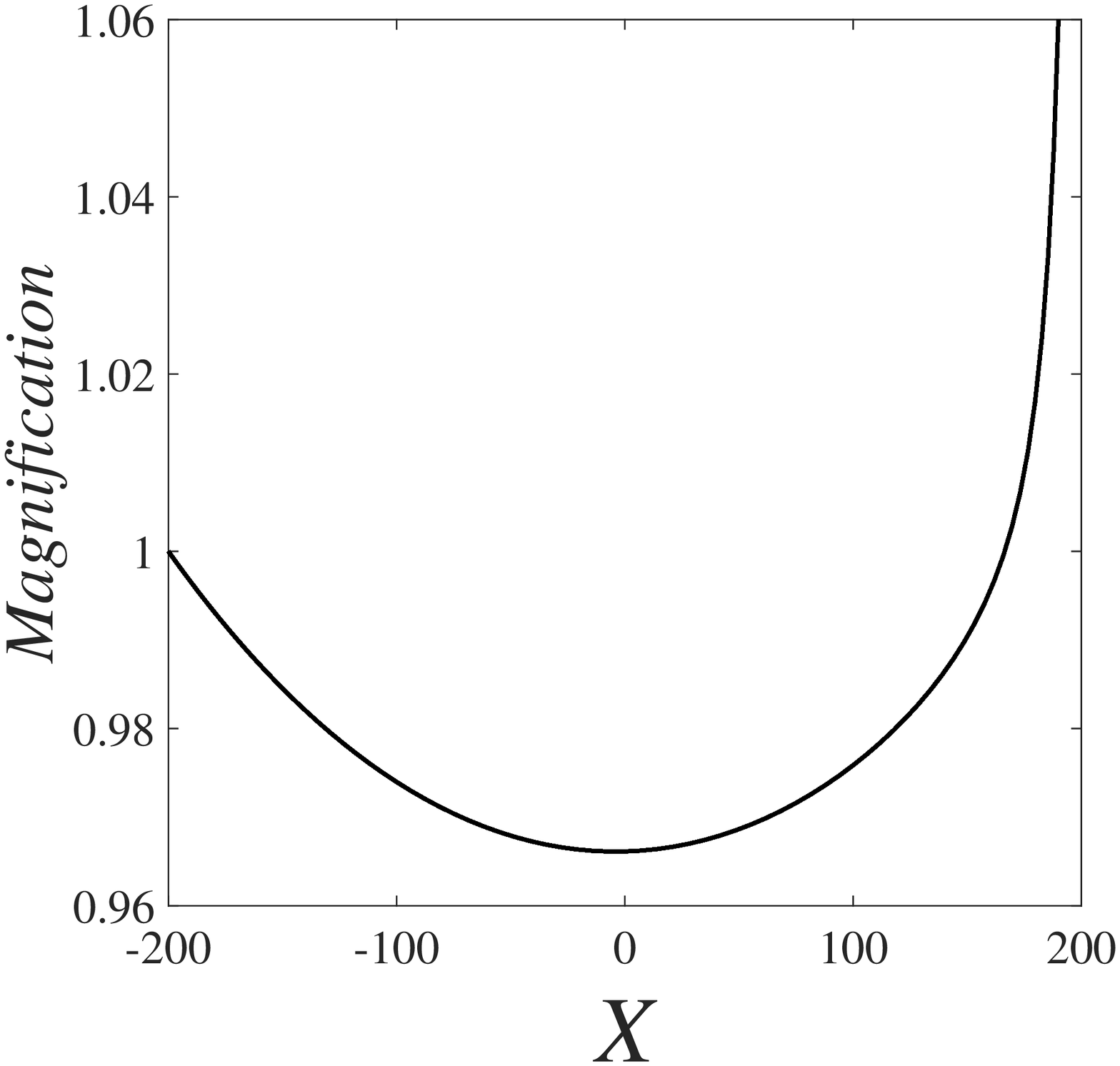}
\includegraphics[width=\columnwidth]{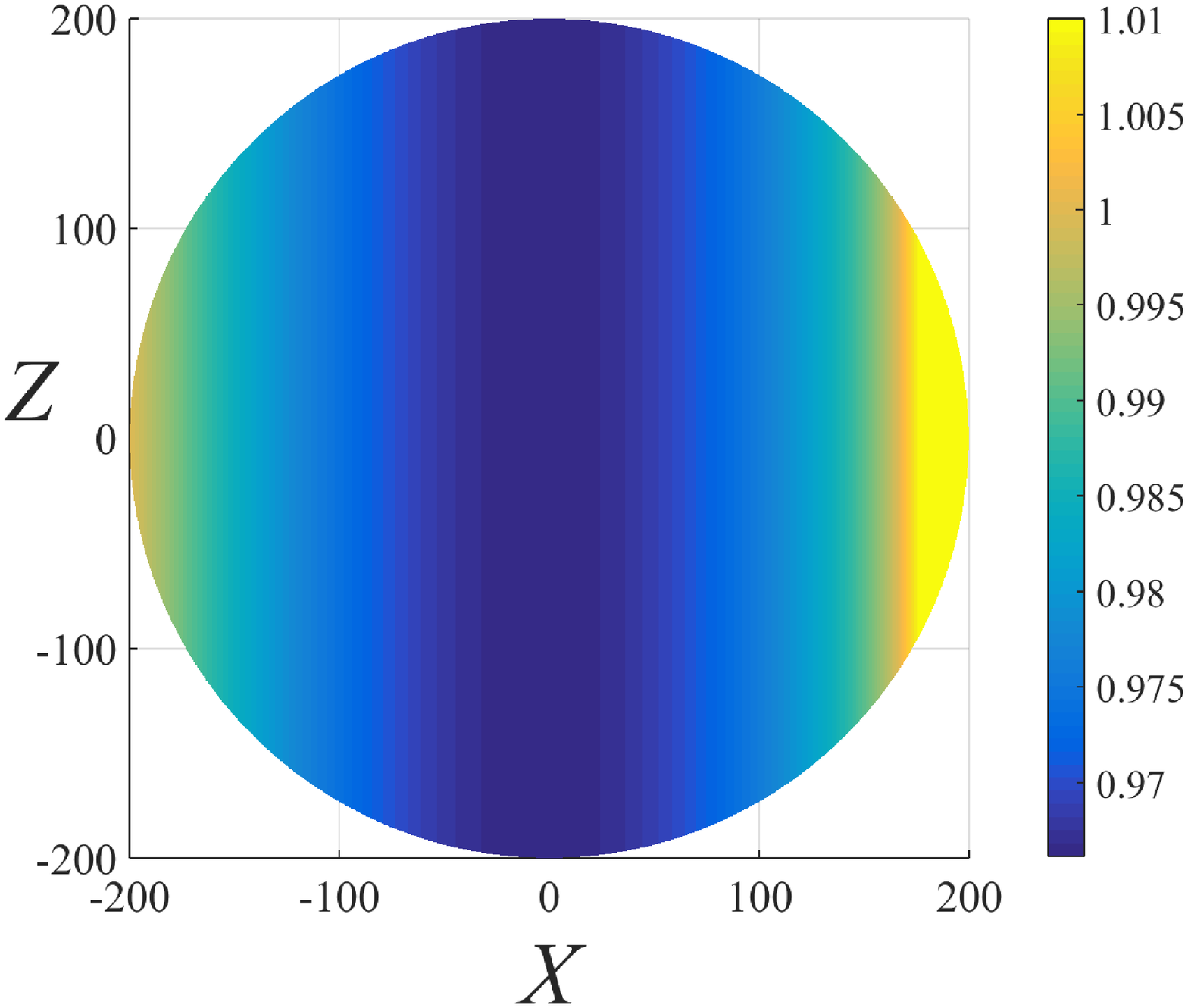}
\label{fig2}
\end{figure}

The complete magnification over the sphere has been plotted in Fig. \ref{fig2}. The system is set up with an observer on the surface of a sphere with radius $200$ centred on the source. There is a lensing object on the $x$-axis at $x=20$ with `mass' $r_s=0.5$. On the left side, magnification is plotted against the observer's $x$-position. It can be seen that directly behind the source (that is, on the opposite side to the lens mass), the magnification is unchanged by the presence of the lens ($M=1$). As the observer moves around the sphere, the source is slightly de-magnified to a minimum value, and then rises arbitrarily high as the observer approaches the (positive) optical axis. A `side on' view of the magnification over the sphere is shown in the right side of the figure. The parameters used are arbitrary and have been chosen to exaggerate the magnification so as to make it clearly visible. In addition, since the magnification on the optical axis is infinite, here it has been truncated to a maximum of $1.01$ in order to increase the level of contrast over the remainder of the sphere. In more realistic systems (with larger distances and a smaller lensing mass) the slight de-magnification, while real, would be so small as to be unmeasurable.

\section{Conclusion and Discussion}
In this note, we have considered the magnification of a source by a lens having the property that it deflects light by an amount which is an approximation of the relativistic deflection of light. We have considered only the primary image. We briefly discussed the thin lens approximation, including the full magnification formula due to such an approach. It was seen that the traditional formula is a small angle approximation to a more general formula. Extending the use of the small angle approximation to large angles gives rise to a paradox which does not exist if the more general result is used. That is, even the thin lens approach need not give rise to the paradox.

In order to account for magnification for any observer position, including observers `behind' the source, a general first order kinematic approach was used to calculate magnification on an observer's sphere. Again, it can be seen that increased magnification near the (positive) optical axis is offset by a slight darkening of the source for the vast majority of observer positions. The full magnification profile is shown in Fig. \ref{fig2}. Interestingly, for observers located on the optical axis behind the source, the distant lensing mass produces no magnification of the primary image, as seen in Equation (\ref{M4}). This differs from the result implied by the use of the moderate angle formula of \citet{wuck}, where the magnification tends to a finite amount always less than one. 

\section*{Acknowledgements}
The authors gratefully acknowledge the financial support for this work provided by Australian Research Council Discovery Grant DP140100094.

\end{document}